\documentclass[aps,preprint,superscriptaddress]{revtex4}

\usepackage{graphicx}

\begin{document}
\bibliographystyle{apsprl}


\title{Low cloud properties influenced by cosmic rays}
\author{Nigel D. Marsh and Henrik Svensmark}
\affiliation{Danish Space Research Institute, Copenhagen, Denmark}
\received{18 May 2000}

\begin{abstract}
The influence of solar variability on climate is currently uncertain. Recent observations have indicated a
possible mechanism via the influence of solar modulated cosmic rays on global cloud cover. Surprisingly the
influence of solar variability is strongest in low clouds ($\leq$ 3km), which points to a microphysical mechanism
involving aerosol formation that is enhanced by ionisation due to cosmic rays. If confirmed it suggests that the
average state of the Heliosphere is important for climate on Earth.
\end{abstract}

\maketitle

\label{intro} The recent discovery that total cloud cover and solar modulated galactic cosmic ray flux (GCR) are
correlated \cite{Svensmark:97,Svensmark:98} suggests that solar variability may be linked to climate variability
through a chain involving the solar wind, GCR and clouds.  The solar wind is a continuous flow of energetic
charged particles (mainly protons and electrons with energies $\sim$KeV) which are released from the sun as a
plasma carrying a fingerprint of the solar magnetic field throughout inter-planetary space.  Influences from the
solar wind are felt at distances well beyond Neptune, possibly up to 200 AU from the sun.  This region of space is
known as the Heliosphere. GCR consists of very energetic particles (mainly protons with typical energies 1 -
20GeV) that originate from stellar processes within our galaxy. Their flux through the solar system is modulated
by the shielding effects of the solar wind whose strength is dependent on the level of solar activity. Those
incident at the Earth are additionally modulated by the geomagnetic field \cite{Lal:67,Herman:78} with cut-off
rigidities of 15 - 0.1GeV from equator to geomagnetic poles. The implication from the observed total cloud cover
- GCR correlation is that climate on Earth could be influenced by the average state of the Heliosphere
(Heliospheric Climate).

Solar forcing of the Earth's climate can be classified into direct and indirect processes. The simplest direct
mechanism is through variations in solar radiative output which is known to vary by 0.1\% over the last solar
cycle, this corresponds to a change of 0.3 W/m$^2$ at the top of the Earth's atmosphere. It is currently believed
that this effect is too small to have had a dominant influence on surface climate, although variations in solar
irradiance may have been larger back in time  \cite{Lean:95}. Indirect effects include solar induced changes in
atmospheric transparency influencing the radiative budget of the planet
 \cite{Svensmark:97,Svensmark:98,Ney:59,Dickinson:75,Pudovkin:95,Tinsley:96}.
 One possibility is that changes in
the solar output of ultra violet (UV) radiation affects temperatures in the stratosphere through absorption by
ozone, which has the potential to influence the large-scale dynamics of the troposphere
\cite{Haigh:96,Shindell:99}.

The observed GCR-cloud correlation introduces another quite different solar influence with the suggestion that
atmospheric ionisation produced by GCR \cite{Svensmark:97,Svensmark:98} affects cloud microphysical properties.
GCR is the dominant source of atmospheric ionisation at altitudes 1-35km over the land and 0-35km over the oceans
with a maximum at $\sim$ 15km due to atmospheric depth. These are regions of the atmosphere in which clouds form.
Clouds are of considerable importance for the Earth's radiation budget, although their exact role is currently
uncertain. Their influence on the vertically integrated radiative properties of the atmosphere result from
cooling through reflection of incoming short wave radiation, and heating through trapping of outgoing long wave
radiation.  The net radiative impact of a particular cloud is mainly dependent upon its height above the surface
and its optical thickness. High optically thin clouds tend to heat while low optically thick clouds tend to cool
\cite{Hartmann:93}. The current climatic estimate for the net forcing of the global cloud cover is $\sim$ 27.7
W/m$^2$ cooling \cite{Hartmann:93,Ramanathan:89,Ardanuy:91}. Thus a significant solar influence on global cloud
properties is potentially important for the Earth's radiation budget
\cite{Svensmark:97,Svensmark:98,Dickinson:75}.  However, the spatial properties of cloud formation vary
considerably. For example, the physics of high ice clouds is quite different to that for low liquid clouds
\cite{Pruppacher:97} thus atmospheric ionisation need not influence all cloud types. It is imperative to
understand which cloud types are influenced by GCR not only from a radiative point of view but, perhaps more
importantly, for identifying a physical mechanism.  Since atmospheric ionisation from GCR reaches a maximum at
high altitudes and latitudes, intuitively, one might expect this is where clouds would feel the greatest effect.
The surprising new result presented here is that only {\it low} cloud properties are varying with GCR. However,
since cloud droplets (in the atmosphere) always condense on an aerosol, this is in agreement with a mechanism
where changes in the atmospheric aerosol distribution influences low liquid clouds. It has recently been shown
that ionisation dominates aerosol production and growth rates when ionisation levels are low and trace gas
concentrations are high such as is found in the lower atmosphere \cite{Yu:00,Yu:00a}.

\label{sec:1} State of the art satellite observations of cloud properties are available as monthly averages from
the International Satellite Cloud Climate Project (ISCCP) D2 analysis derived from the Top Of Atmosphere (TOA)
radiance for the period July 1983 to September 1994 \cite{Schiffer:85,Rossow:91,Rossow:96}. Infrared (IR)
measurements (uncertainty 1-2K \cite{Brest:97}) are preferred due to their superior spatial and temporal
homogeneity over visual observations that can only be detected during daylight.  Cloud cover is obtained from an
algorithm using the TOA IR statistics to identify the cloudiness  on an equal area grid (280km x 280km). Cloud
top temperatures (CT) and pressures (CP) are obtained from an ISCCP IR model constrained by water vapour and
vertical temperature profiles retrieved from the TIROS Observed Vertical Sounder (TOVS) \cite{Rossow:96}. CT and
CP are found by assuming an opaque blackbody cloud, and adjusting the cloud's pressure level (effectively cloud
height) in the model until the reconstructed outgoing IR flux at TOA matches that observed. Based on retrieved
CP, clouds are divided into Low $> 680$hPa ($ <3.2$km), Middle $=680-440$hPa ($3.2-6.5$km), and High $<440$hPa
($>6.5$km).

Figure 1 indicates that a 2-3 \% change in low cloud cover correlates with GCR over the whole period, while the
middle and high clouds do not (uncertainties in cloud cover $\leq 1\%$ \cite{Rossow:95}).  The spatial
distribution of this low cloud cover correlation is shown in Fig. 2a. Regions displaying a correlation $r \geq
0.6$ cover a highly significant 15.8 \% fraction of the Earth surface - see Fig. 2 caption. The probability of
obtaining such a surface fraction by chance was found to be better than $10^{-3}$ from an ensemble of Monte Carlo
simulations. Each member of the ensemble consisted of $N$ independent artificial cloud time series, where $N$
($\sim 160$) was the spatial degree of freedom determined from spatial cloud correlations. The most restrictive
test was by generating the artificial cloud series from a fourier transform of the real cloud data, randomizing
the phases, and fourier transforming back. Note that the high correlation in Fig. 1c, where $r=0.63$ and $r=0.92$
for the 12 month running mean (confidence limits assuming t-distribution $< 10^{-5}$), is obtained by taking the
global average of cloud anomalies used in Fig. 2a which reduces fluctuations due to both instrument noise and
internal climate variability.

However, at these time scales GCR ionisation is not the only mechanism affecting low clouds, there are of course
many other decadal processes in the climate system which are important.  The small differences in leads and lags
are close to the satellites resolution and one should not expect a perfect correlation. What is surprising is
that despite these limitations a signal of solar variability in low cloud cover is dominant at time-scales longer
than 1 year. Svensmark \cite{Svensmark:98} argued that there is a better agreement with GCR rather than solar
irradiance for total cloud cover.  This is also true for the {\it low cloud cover} in Figure 1c, which suggests
that low cloud cover is responding to cosmic ray ionisation in the atmosphere rather than direct changes in solar
irradiance.

Currently satellites cannot detect multi-layer cloud, thus high and middle clouds can obscure clouds below.  From
this point of view low clouds contain the least contaminated signal giving greater confidence to this result.
However, if a cloud is transmissive then the satellite observes both radiation from below the cloud and radiation
from the cloud itself. Since ISCCP defines all clouds to be opaque, the CT of transmissive clouds is
overestimated such that their altitude appears lower in the IR model than in reality. For the case of
transmissive clouds CT represents a weighted average based on emissivity of the clouds present in a column
scene.  However, the long term global trend in low clouds is not explained by an artifact due to mixing with
clouds from above since no GCR signal is apparent in the middle and high clouds over the period of observations
(Figures 1a and b). But low clouds could be contaminated with overlaying very thin undetected transmissive cloud,
e.g., high thin cirrus, and the signal of solar variability could be due to undetected high cloud.  This is
perhaps more intuitive since GCR atmopsheric ionisation is greater at higher altitudes, and stratospheric heating
due to UV possesses a strong solar signal \cite{Labitzke:99}.  However, it will be shown that this is not the
case.

Although the ISCCP analysis poorly detects high very thin cloud, a comparison with HIRS (High resolution Infrared
Radiation Sounder) measurements suggests that ISCCP captures the general trends of high thin cloud
 \cite{Jin:96}. If a solar signal does exist in high cloud, for whatever reason, one would expect to see a signal in
those high clouds that are detected by ISCCP, Figure 1a.  However, no such signal is observed, thus there are
good reasons to believe that the long term trends in low cloud cover are due to real low clouds responding to GCR.

The low cloud top temperature parameter also correlates with GCR over large regions of the Earth. Figure 2b
reveals a band of significantly high correlation centered around the tropics, while there are no significant
correlations for middle and high cloud top temperatures (not shown).  The ISCCP IR statistics cannot easily
distinguish very low cloud top temperatures, which are relatively warm, from surface temperatures.  Thus the
modelled surface temperatures, ST, will be contaminated with temperatures from very low cloud. It is interesting
to note that ST contains a very similar GCR correlation map (not shown) as that for low cloud CT in Fig. 2b. The
lack of correlation at high latitudes in Fig. 2b is currently not understood, but may be a feature of a possible
GCR-cloud mechanism outlined below.

\label{sec:2} The opaque cloud assumption in the ISCCP IR model excludes micro-physical properties and so
constrains cloud variability to appear only in cloud 'model height', thus introducing an element of artificial
variability into CT. Observed properties of low level maritime clouds suggests that they are not opaque
\cite{Heymsfield:93}. Relaxing the opaque assumption allows for cloud variability to additionally manifest itself
through changes in cloud optical density. Cloud optical density depends on processes affecting the cloud droplet
size distribution, and cloud vertical extent. Since all atmospheric liquid water droplets form on cloud
condensation nuclei (CCN), the droplet size distribution depends on the density of atmospheric aerosols activated
as CCN, while cloud thickness is influenced by atmospheric vertical temperature profiles. The abundance of CCN is
determined by both the level of supersaturation and the number of aerosols present in the atmosphere able to act
as CCN. Increases in supersaturation, typically between $0.1 \%$ and a few percent, activates increasingly
smaller aerosols. A solar signal could enter low cloud properties through influencing: atmospheric vertical
temperature profiles, water vapour, or aerosol to CCN activation processes.  In the following it is argued that
the latter is a more likely explanation.

Thermodynamic properties of the atmosphere where low clouds form are affected via changes to tropospheric
circulation. Studies with general circulation models have indicated that solar induced variability in the
stratosphere can influence the vertical circulation of the troposphere \cite{Haigh:96,Shindell:99}. However, TOVS
observations of the vertical profiles of water vapour and temperature demonstrate little correlation with GCR.
This suggests that the influence of variability in solar irradiance on local thermodynamic properties in the
atmosphere is not responsible for the observed changes in low cloud properties. This might not be surprising
given that variability in solar irradiance agrees poorly with changes in low cloud properties
 \cite{Svensmark:98}.

Assuming typical atmospheric water vapour saturation, the abundance of CCN is determined through properties of
the background aerosol size distribution ($\sim 0.01-1.0$ $\mu$m). Production of aerosol can be due to many
processes involving: gas-particle conversion, droplet-particle conversion, i.e., evaporation of water droplets
containing dissolved matter, and bulk particles from the surface, e.g., smoke, dust, or pollen
\cite{Pruppacher:97}.  Observations of spectra in regions of low cloud formation indicate that aerosols are
produced locally.  In the troposphere it has been suggested that ionisation contributes to the gas-particle
formation of ultrafine ($<0.02$ $\mu$m) aerosol.  Model studies indicate that this process could contribute a
stable concentration of several hundred particles per cm$^{3}$ at sizes $>0.02$ $\mu$m \cite{Turco:98}. This is
comparable to the total number of condensation nuclei in maritime air ($\sim 100$ cm$^{-3}$) \cite{Pruppacher:97}.
Observations of aerosol growth into the aged aerosol distributions generating CCN have been interpreted to be
influenced by the presence of ionisation \cite{Yu:00,Turco:98,Horrak:98}.  A recent study of ion mediated
nucleation by Yu and Turco \cite{Yu:00a} indicates that the nucleation rate of ultrafine aerosol is generally
limited by ionisation from GCRs in the lower maritime atmosphere.  In contrast, they show that nucleation in the
upper atmosphere is limited by concentrations of trace gases, e.g., H$_{2}$SO$_{4}$.  Although it is currently
uncertain how the ultrafine aerosol evolves into CCN it could explain why only low cloud properties are
responding to GCR modulation.  It is less clear why this modulation should be restricted to lower latitudes, seen
particularly in low cloud top temperatures (Fig. 2b), which appears to contradict a larger geomagnetic shielding
of cosmic rays at the equator \cite{Svensmark:97}. However, ion mediated nucleation saturates when levels of
ionisation are high relative to concentrations of trace gases \cite{Yu:00a}, so a latitudinal dependence of
either or both of these could be involved. This is currently an area of further research.

\label{sec:3}

Based on the ISCCP D2 IR cloud data there is a clear correlation between GCR and properties of low clouds in
contrast to middle and high clouds. Since the correlation is seen both in low cloud cover and low cloud top
temperature, the case for solar induced variability of low clouds is strengthened. Observations of atmospheric
parameters from TOVS do not support a solar-cloud mechanism through tropospheric dynamics influenced by UV
absorption in the stratosphere. Instead, it is argued that a mechanism involving solar modulated GCR is possible.
It has been speculated for some time that ionisation is important for aerosol production and growth in the
troposphere.  Recent studies indicate that ionisation is a limiting process for aerosol nucleation in the lower
maritime atmosphere, thus it is not unreasonable to imagine that systematic variations in GCR ionisation could
affect atmospheric aerosols acting as CCN and hence low cloud properties. If such mechanisms can be confirmed the
implications for clouds and climate are far reaching, and suggests that Heliospheric climate can influence
climate on Earth. Based on observations, Lockwood {\it et al.} have shown that since 1964 the strength of the
solar magnetic flux, shielding the Earth from GCR, has increased by 41\% while GCR has decreased by 3.7\%
\cite{Lockwood:99}. Further, they claim that the solar magnetic flux has more than doubled over the last
century.  Based on this doubling and assuming a GCR - Low cloud mechanism exists, a crude estimate for the
century trend in low cloud radiative forcing is a warming of $1.4$ Wm$^{-2}$ \cite{Marsh:00}. Thus, if there is a
systematic variation in low cloud properties caused by solar variability it could have important implications for
the evolution of Earth's climate.



\begin{figure}[h]
\label{fig:1}
  \includegraphics[]{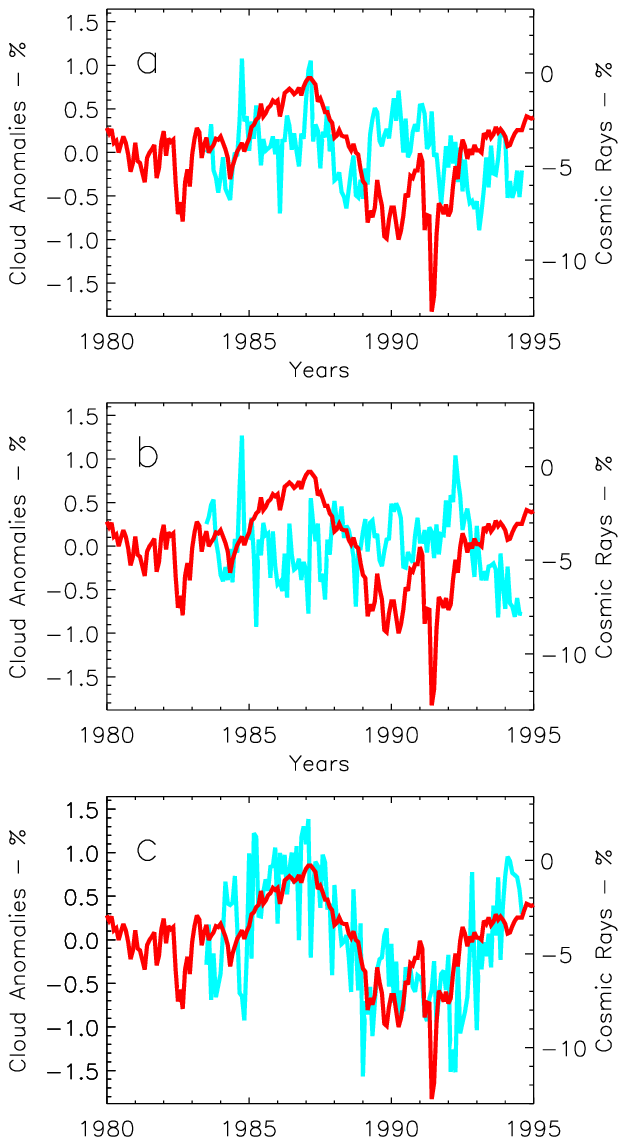}
  \caption{Global average of monthly cloud anomalies for a) high
    ($<440$ hPa), b) middle ($440-680$ hPa), and c) low ($>680$ hPa) cloud
    cover (blue). To compute the monthly cloud anomalies the annual cycle is
    removed by subtracting the climatic monthly
    average (July 1983 - June 1994) from each month on an equal area grid before
    averaging over the globe.  The
    global average of the annual cycle over this period for high, middle and
    low IR detected clouds is 13.5\%, 19.9\%, and 28.7\% respectively.
    The cosmic rays (red) represent neutron counts observed
    at Huancayo (cut-off rigidity 12.91 GeV) and normalised to Oct 1965.
    }
\end{figure}

\begin{figure}[h]
\label{fig:2}\includegraphics[]{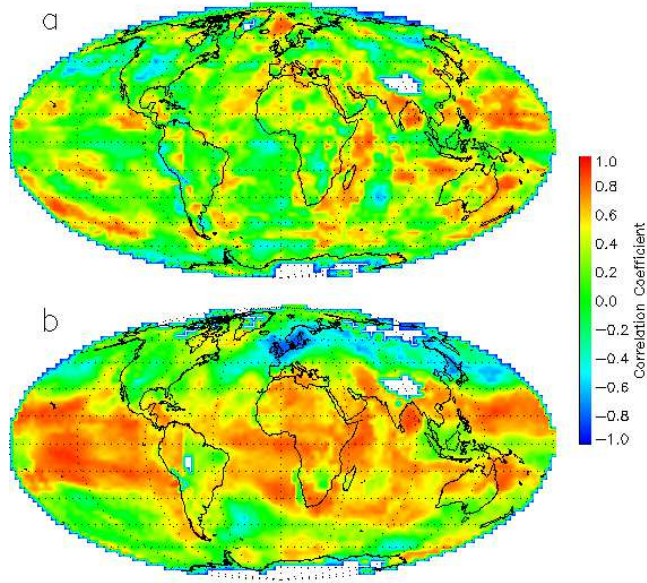}
 \caption{Global correlation maps of GCR with anomalies of a) Low IR
    cloud cover, and b) Low IR cloud top temperature (CT).  The low
    IR cloud cover is calculated as in Fig. 1c, while the low
    cloud CT are obtained from the ISCCP IR
    model.  White pixels indicate regions with either no data or an
    incomplete monthly time series. The correlation coefficients, $r$, are
    calculated from the 12 month running mean at each grid point. Regions
    of the Earth with $r \geq 0.6$ are a) 15.8\%, and
    b) 34.6\%. The probability of obtaining these surface fractions by chance is better than $10^{-3}$.}

\end{figure}

\end{document}